\documentclass[epsfig,12pt]{article}
\usepackage{makeidx}
\usepackage{amsmath}
\usepackage{amsfonts}
\usepackage{amssymb}
\usepackage{graphicx}

\input epsf.sty
\makeatletter \@addtoreset{equation}{section}
\renewcommand{\theequation}{\thesection.\arabic{equation}}
\input{tcilatex}
\begin{document}

\title{\rightline{\mbox{\small
{Inanotech/LMS/09/04}}} \textbf{Electronic Properties and Hidden Symmetries
of Graphene}}
\author{L.B Drissi$^{{\scriptsize 1}}$ {\small \thanks{%
b.drissi@inanotech.ma} \ }, E.H Saidi$^{{\scriptsize 1,2}}$ {\small \thanks{%
h-saidi@fsr.ac.ma} \ }, M. Bousmina$^{{\scriptsize 1}}$ \\
{\small 1. INANOTECH, Institute of Nanomaterials and Nanotechnology, Rabat,
Morocco,}\\
{\small 2. Lab/UFR-Physique des Hautes Energies, Facult\'{e} des Sciences,
Rabat, Morocco.}}
\maketitle

\begin{abstract}
Using the relation between the structural and the electronic properties of
honeycomb, we study the hidden $SU\left( 3\right) $ symmetry of the graphene
monolayer and exhibit the link with its electronic properties. We show that
the conservation law of incoming and outgoing electronic momenta at each
site of graphene is solved in terms of SU$\left( 3\right) $ representations;
and the Fourier waves $\tilde{\phi}\left( k_{x},k_{y}\right) $ of the
hopping electron may be classified by SU$\left( 3\right) $ highest weight
multiplets $\phi _{p,q}\left( \xi \right) $. It is also shown that the
phases $\arctan \frac{k_{y}}{k_{x}}$ of the waves are quantized as $\frac{%
\left( p+q\right) }{\left( p-q\right) }\sqrt{3}$ with $p,$ $q$ positive
integers. Other features are also discussed.
\end{abstract}


\section{Introduction}

\begin{figure}[tbph]
\begin{center}
\hspace{0cm} \includegraphics[width=10cm]{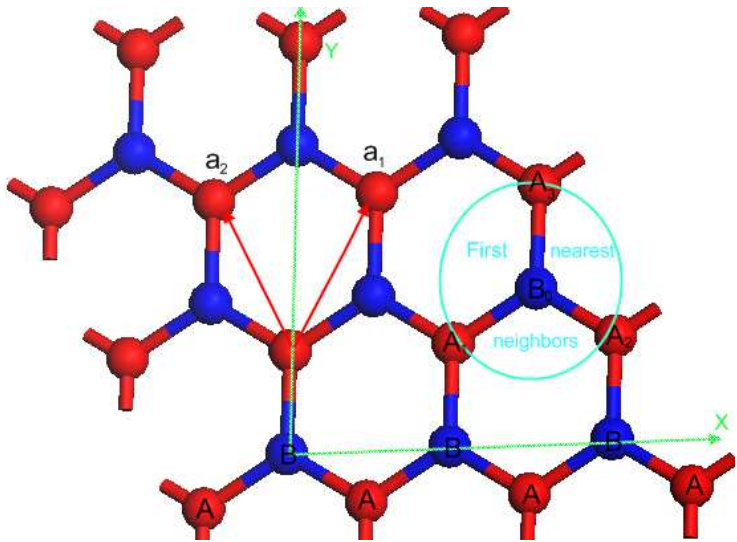}
\end{center}
\par
\vspace{-0.5cm}
\caption{{\protect\small Sublattices }$\mathcal{A}${\protect\small \ and }$%
\mathcal{B}${\protect\small \ of the honeycomb. A-type carbons are given by
red balls and B-type atoms by blue ones. Each carbon has \emph{3} first
nearest neighbors and \emph{6} second nearest ones.}}
\label{1}
\end{figure}

Graphene is a system of carbon atoms in the sp$^{2}$\ hybridization forming
a 2D honeycomb lattice. This is a planar system made of two triangular
sublattices A and B and constitute the building block of graphite. Since its
experimental realization in 2004, the study of the electronic properties of
graphene with and without external fields has been a big subject of interest 
\textrm{\cite{1}-\cite{4}}; some of its main physical aspects were recently
reviewed in \textrm{\cite{5} }and refs therein. This big interest into
graphene and its derivatives is because they offer a real alternative for
silicon based technology\textrm{\ }and bring together issues from condensed
matter and high energy physics \textrm{\cite{6}-\cite{12} }allowing a better
understanding of the electronic band structure as well as their special
properties\textrm{.} \newline
In this paper, we focus on an unexplored issue of 2D graphene by studying
the link between specific electronic properties and a class of hidden
symmetries of the honeycomb. These symmetries allow to get more insight in
the transport property of the electronic wave modes and may be used to
approach the defects and boundaries \textrm{\cite{13}}. The existence of
these hidden symmetries; in particular of the remarkable hidden $SU\left(
3\right) $ invariance considered in this study, may be motivated from
several views. For instance from the structure of the first nearest carbon
neighbors like for the typical $\left\langle A_{0}\text{-}B_{1}\right\rangle 
$, $\left\langle A_{0}\text{-}B_{2}\right\rangle ,$ $\left\langle A_{0}\text{%
-}B_{3}\right\rangle $ see fig(\ref{1}) for illustration and more details. 
These are basic patterns generating three $SU\left( 2\right) $ symmetries
contained in the hidden SU$\left( 3\right) $ invariance of honeycomb. The $%
\left\langle A\text{-}B\right\rangle $ patterns transform in the isospin $%
\frac{1}{2}$ representations and describe the electronic wave doublets $\phi
_{{\scriptsize \pm }\frac{{\scriptsize 1}}{{\scriptsize 2}}}=\left[ a\left( 
\mathbf{r}\right) ,b\left( \mathbf{r}\right) \right] $ interpreted as
quasi-relativistic 2D spinors in the nearby of the Dirac points \textrm{\cite%
{5}}. The $SU\left( 3\right) $ hidden symmetry of honeycomb is also encoded
in the second nearest neighbors $\left\langle \left\langle A_{0}\text{-}%
A_{i}\right\rangle \right\rangle $ and $\left\langle \left\langle B_{0}\text{%
-}B_{i}\right\rangle \right\rangle $, $i=1,...,6$, which capture information
on its adjoint representation where the six $\left\langle \left\langle A_{0}%
\text{-}A_{i}\right\rangle \right\rangle $ (and similarly for $\left\langle
\left\langle B_{0}\text{-}B_{i}\right\rangle \right\rangle $) are associated
with the six roots of $SU\left( 3\right) $. In addition to above mentioned
properties, hidden symmetries of graphene are also present in the framework
of the tight binding model with hamiltonian, 
\begin{equation*}
\begin{tabular}{lll}
$H=$ & $-t_{1}\sum\limits_{r_{i}\in A}\sum\limits_{n=1}^{3}a\left( \mathbf{r}%
_{i}\right) b^{\dagger }\left( \mathbf{r}_{i}{\scriptsize +}\mathbf{\delta }%
_{n}\right) +hc$ &  \\ 
& $-t_{2}\sum\limits_{\left\langle \left\langle r_{i},r_{j}\right\rangle
\right\rangle }a\left( \mathbf{r}_{i}\right) a^{\dagger }\left( \mathbf{r}%
_{j}\right) +b\left( \mathbf{r}_{i}\right) b^{\dagger }\left( \mathbf{r}%
_{j}\right) $ & $,$%
\end{tabular}%
\end{equation*}%
where the fermionic creation and annihilation operators $a,$ $a^{\dagger },$ 
$b,$ $b^{\dagger }$ are respectively associated to the pi-electrons of each
atom of the sublattices A and B and where the three \emph{relative} vectors $%
\mathbf{\delta }_{1}$, $\mathbf{\delta }_{2}$, $\mathbf{\delta }_{3}$ define
the first nearest neighbors as depicted in fig(\ref{1}). These 2D vectors
are globally defined on the honeycomb and obey the remarkable constraint
equation $\mathbf{\delta }_{1}+\mathbf{\delta }_{2}+\mathbf{\delta }_{3}=%
\mathbf{0}$ which, a priori, encodes also information on the electronic
properties of graphene. Throughout this study, we show amongst others, that
the three above mentioned $SU\left( 2\right) $'s are intimately related with
these $\mathbf{\delta }_{n}$'s which, as we will see, are nothing but roots
of $SU\left( 3\right) $. We also show that the mapping of the condition $%
\sum_{n=1}^{3}\mathbf{\delta }_{n}=\mathbf{0}$ to the momentum space can be
interpreted as a condition on the conservation of total momenta at each site
of honeycomb whose solutions are classified by highest weight state
representations of the $SU\left( 3\right) $ symmetry. We show moreover that
the hamiltonian $H$ of the tight binding model for first nearest neighbors
has an interpretation in terms of the $F^{\pm \beta }$ step operators of
these $SU\left( 2\right) $'s opening a window for more insight into the
study of the electronic correlations in graphene and cousin systems. \newline
The organization of this paper is as follows: In section 2, we exhibit the $%
SU\left( 3\right) $ symmetry of graphene. In section 3, we give a field
theoretic interpretation of the geometric constraint equation $\mathbf{%
\delta }_{1}+\mathbf{\delta }_{2}+\mathbf{\delta }_{3}=\mathbf{0}$ both in
real and reciprocal honeycomb. We also use the roots and weights of hidden $%
SU\left( 3\right) $ symmetry to study aspects of the electronic properties
of graphene. In section 4, we develop the relation between the energy
dispersion and the hidden symmetries. In section 5, we give the conclusion
and a perspective.

\section{ Hidden symmetries of graphene}

\begin{figure}[tbph]
\begin{center}
\hspace{0cm} \includegraphics[width=14cm]{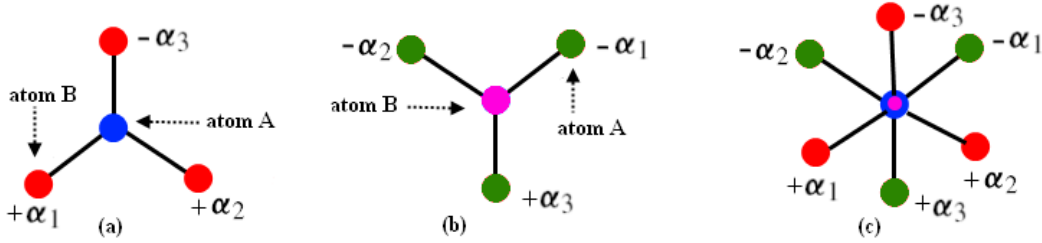}
\end{center}
\par
\vspace{-0.5 cm}
\caption{{\protect\small \ (a) Nearest neighbors of a A- type atom. (b)
Nearest neighbors of a B-type atom. (c) The fusion of the two vertices gives
the adjoint representation of SU}$\left( {\protect\small 3}\right) $%
{\protect\small .}}
\label{2}
\end{figure}

In dealing with ideal 2D graphene, one notices the existence of a hidden $%
SU\left( 3\right) $ group symmetry underlying the crystallographic structure
of the honeycomb lattice and governing the hopping of the pi-electrons
between the closed neighboring carbons. To exhibit this hidden $SU\left(
3\right) $ symmetry, let us start by examining some remarkable features on
the graphene lattice and show how they are closely related to SU$\left(
3\right) $. \newline
Refereing to the two sublattices of the graphene monolayer by the usual
letters A and B generated by the vectors $\mathbf{a}_{1,2}=\frac{a}{2}(\pm 
\sqrt{3},3)$ and the relative ones $\mathbf{\delta }_{1,2}=\frac{a}{2}\left(
\pm \sqrt{3},1\right) $, and denoting by $\phi _{A}\left( \mathbf{r}%
_{i}\right) $ and $\phi _{B}\left( \mathbf{r}_{j}\right) $ the wave
functions of the corresponding pi-electrons, one notes that the interactions
between the first nearest atoms involve two kinds of trivalent vertices
capturing data on $SU\left( 3\right) $ symmetry, see fig(\ref{2}) for
illustration. 
This hidden $SU\left( 3\right) $ invariance can be made more explicit by
remarking that the relative vectors $\mathbf{\delta }_{1},$ $\mathbf{\delta }%
_{2}$ and $\mathbf{\delta }_{3}=-\mathbf{\delta }_{1}-\mathbf{\delta }_{2}$
describing the three first closed neighbors to a A- type carbon at site $%
\mathbf{r}_{i}$ of the honeycomb together with their opposites $-\mathbf{%
\delta }_{n}$ for B-type carbons are precisely the roots of the $SU\left(
3\right) $ algebra. Indeed, if forgetting about the scale dimension and
thinking about the carbon-carbon distance $a\simeq 1.42$ $A^{%
{{}^\circ}%
}$ as the irrational number\textrm{\footnote{%
Although it isn't a necessary condition for our analysis, this number is
mysterious as it corresponds precisely to the length $\left\Vert \mathbf{%
\alpha }_{i}\right\Vert ^{2}=2$ of the roots of simply laced Lie algebras;
see also end of conclusion.}} $a=\sqrt{2}$, one gets the remarkable
identification 
\begin{equation}
\begin{tabular}{llllll}
$\mathbf{\delta }_{1}=\mathbf{\alpha }_{1}$ & , & $\mathbf{\delta }_{2}=%
\mathbf{\alpha }_{2}$ & , & $\mathbf{\delta }_{3}=-\mathbf{\alpha }_{3}$ & ,%
\end{tabular}
\label{r}
\end{equation}%
with $\mathbf{\alpha }_{1}$\ and $\mathbf{\alpha }_{2}$ being the two simple
roots of $SU\left( 3\right) $. This symmetry can be also exhibited by
computing the intersection matrix $\mathbf{\delta }_{i}\cdot \mathbf{\delta }%
_{j}$ of the two generators $\mathbf{\delta }_{1}$ and $\mathbf{\delta }_{2}$%
, 
\begin{equation}
\mathbf{\delta }_{i}\cdot \mathbf{\delta }_{j}=\frac{a^{2}}{2}\left( 
\begin{array}{cc}
2 & -1 \\ 
-1 & 2%
\end{array}%
\right) ,
\end{equation}%
which turns out to be proportional to the Cartan matrix of the $SU\left(
3\right) $ algebra, $A_{ij}=\mathbf{\alpha }_{i}\cdot \mathbf{\alpha }_{j}$.
Recall that the Lie algebra of SU$\left( 3\right) $ has rank two, eight
generators $\left\{ F^{a}\right\} $ and commutation relations that read in
the Gell-Mann basis like 
\begin{equation}
\left[ F^{a},F^{b}\right] =if^{abc}F^{c},  \label{g}
\end{equation}%
with antisymmetric structure constants as $f_{123}=1$, $%
f^{147}=f^{516}=f^{246}=f^{257}=f^{345}=f^{637}=\frac{1}{2}$ and $%
f^{458}=f^{678}=1$. In the Cartan-Weyl basis, useful for physical
interpretations, we take $h^{1}=\sqrt{2}F^{8}$ and $h^{2}=\sqrt{2}F^{3}$ as
the two Cartan terms and the six step operators like $\sqrt{2}U_{3}^{\pm
}=F^{1}\pm iF^{2}$, $\sqrt{2}U_{2}^{\pm }=F^{4}\pm iF^{5}$, $\sqrt{2}%
U_{1}^{\pm }=F^{6}\pm iF^{7}$. The new commutation relations following from (%
\ref{g}) read as follows,%
\begin{equation}
\begin{tabular}{ll}
$\left[ h^{i},U_{n}^{\pm }\right] =\pm \delta _{n}^{i}$ $U_{n}^{\pm }$ & ,%
\end{tabular}%
\end{equation}%
with $i=1,2$ and where the $\mathbf{\delta }_{n}$'s are the same vectors as
in the graphene. Notice in passing that the hidden SU$\left( 3\right) $
described above seems to be just a sub-symmetry of a larger one since the $%
\mathbf{\delta }_{n}$'s obey the constraint relation 
\begin{equation}
\sum_{n=1}^{3}\mathbf{\delta }_{n}=\mathbf{0.}  \label{de}
\end{equation}%
which might hide an affine $S\hat{U}\left( 3\right) $ Kac-Moody symmetry 
\textrm{\cite{14,15} }since\textrm{\ }the intersection matrix of the three
relative $\mathbf{\delta }_{i}$- vectors reads as $\mathbf{\delta }_{i}\cdot 
\mathbf{\delta }_{j}=\frac{a^{2}}{2}\hat{A}_{ij}$ with%
\begin{equation}
\begin{tabular}{ll}
$\hat{A}_{ij}=\left( 
\begin{array}{ccc}
2 & -1 & -1 \\ 
-1 & 2 & -1 \\ 
-1 & -1 & 2%
\end{array}%
\right) $ & ,%
\end{tabular}%
\end{equation}%
describing exactly the generalized Cartan matrix of affine $S\hat{U}\left(
3\right) $. Below, we shall restrict our study to the hidden ordinary $%
SU\left( 3\right) $ symmetry of the graphene\ and think about (\ref{de}) as
a physical constraint equation governing the electronic properties of the
graphene.

\section{Electronic properties and $SU\left( 3\right) $ symmetry}

Quantum mechanically, there are two approaches to deal with the geometrical
constraint relation (\ref{de}). The first one is to work in real space and
think about it as the conservation law of total space-time probability
current densities at each site $\mathbf{r}_{i}$ of the honeycomb. The second
approach relies on moving to the dual space where this constraint relation
and the induced electronic properties get a remarkable interpretation in
terms of $SU\left( 3\right) $ representations.

\subsection{Conservation of total current density}

In the real space, the way we interpret eq(\ref{de}) is in terms of the\
relation between the time variation of the probability density $\rho \left(
t,\mathbf{r}_{i}\right) =\left\vert \phi \left( t,\mathbf{r}_{i}\right)
\right\vert ^{2}$ of the electron at site $\mathbf{r}_{i}$ and the sum 
\begin{equation}
\sum_{n=1}^{3}\mathbf{J}_{{\scriptsize \delta }_{n}}\left( t,\mathbf{r}%
_{i}\right) =\mathbf{J}\left( t,\mathbf{r}_{i}\right) \text{ \ \ \ ,}
\end{equation}%
of incoming and outgoing probability current densities along the $\delta
_{n} $- directions. On one hand, because of the \emph{equiprobability} in
hopping from the carbon at $\mathbf{r}_{i}$ to each one of the three nearest
carbons at $\mathbf{r}_{i}+\mathbf{\delta }_{n}$, the norm of the $\mathbf{J}%
_{{\scriptsize \delta }_{n}}$- vector current densities should be equal and
so they should have the form 
\begin{equation}
\begin{tabular}{llll}
$\mathbf{J}_{{\scriptsize \delta }_{n}}\left( t,\mathbf{r}_{i}\right)
=j\left( t,\mathbf{r}_{i}\right) \mathbf{e}_{n}$ & , & $n=1,2,3$ & .%
\end{tabular}%
\end{equation}%
These probability current densities together with the unit vectors $\mathbf{e%
}_{n}=\frac{\mathbf{\delta }_{n}}{a}$ point in the different $\mathbf{\delta 
}_{n}$- direction; but have the same non zero norm: $\left\Vert \mathbf{J}_{%
{\scriptsize \delta }_{1}}\right\Vert =\left\Vert \mathbf{J}_{{\scriptsize %
\delta }_{2}}\right\Vert =\left\Vert \mathbf{J}_{{\scriptsize \delta }%
_{3}}\right\Vert =\left\vert j\right\vert $. Substituting in the above
relation, the total probability current density $\mathbf{J}\left( t,\mathbf{r%
}_{i}\right) $ at the site $\mathbf{r}$ and time $t$ takes then the
factorized form 
\begin{equation}
\mathbf{J}\left( t,\mathbf{r}\right) =\frac{j\left( t,r\right) }{a}\left(
\sum_{n}\mathbf{\delta }_{n}\right) .  \label{tot}
\end{equation}%
On the other hand, by using the\ Schrodinger equation $i\hbar \frac{\partial
\phi }{\partial t}=\left( -\frac{\hbar ^{2}}{2m}\nabla ^{2}+V\right) \phi $
describing the interacting dynamics of the electronic wave at $\mathbf{r}$,
we have the usual conservation equation,%
\begin{equation}
\frac{\partial \rho \left( t,\mathbf{r}\right) }{\partial t}+\func{div}%
\mathbf{J}\left( t,\mathbf{r}\right) =0\text{ \ \ },
\end{equation}%
with probability density $\rho \left( t,\mathbf{r}\right) $ as before and $J=%
\frac{i\hbar }{2m}\left( \phi \nabla \phi ^{\ast }-\phi ^{\ast }\nabla \phi
\right) $ with m the mass of the electron and $\phi =\phi \left( t,\mathbf{r}%
\right) $ its wave. Moreover, assuming $\frac{\partial \rho }{\partial t}=0$
corresponding to stationary electronic waves $\phi \left( t,\mathbf{r}%
\right) =e^{i\omega t}\phi \left( \mathbf{r}\right) $, it follows that the
space divergence of the total current density vanishes identically; $\func{%
div}\mathbf{J}=0$. This constraint equation shows that generally $\mathbf{J}$
should be a curl vector; but physical consideration indicates that we must
have 
\begin{equation}
\mathbf{J}\left( t,\mathbf{r}\right) =0\text{ \ \ ,}
\end{equation}%
in agreement with Gauss-Stokes theorem $\int_{\mathcal{V}}\func{div}\mathbf{J%
}$ $d\mathcal{V}$ $\mathbf{=}$ $\int_{\partial \mathcal{V}}\mathbf{J.}d%
\mathbf{\sigma }$ leading to the same conclusion. Combining the property $%
\mathbf{J}\left( t,\mathbf{r}\right) =0$ with its factorized expression $%
\frac{j}{a}\left( \sum_{n}\mathbf{\delta }_{n}\right) $ given by eq(\ref{tot}%
) together with $j\neq 0$, we end with the constraint relation $\sum_{n}%
\mathbf{\delta }_{n}=0$.\ 

\subsection{Conservation of total phase}

In the dual space of the electronic wave of graphene, the constraint
relation (\ref{de}) may be interpreted in two different, but equivalent,
ways; first in terms of the conservation of the total relative phase $\Delta
\varphi _{{\scriptsize tot}}=\sum \mathbf{k.}\Delta \mathbf{r}$ of the
electronic waves induced by the hopping to the nearest neighbors. The second
way is in terms of the conservation of the total momenta at each site of the
honeycomb. \newline
Decomposing the wave function $\phi \left( \mathbf{r}\right) $, associated
with a A-type carbon at site $\mathbf{r}$, in Fourier modes as $%
\sum_{k}e^{i2\pi \mathbf{k}\cdot \mathbf{r}}$ $\tilde{\phi}\left( \mathbf{k}%
\right) $; and similarly for the B-type neighboring ones $\phi \left( 
\mathbf{r}+\mathbf{\delta }_{n}\right) =\sum_{k}e^{i2\pi \mathbf{k}\cdot 
\mathbf{r}}$ $\tilde{\phi}_{n}\left( \mathbf{k}\right) $ with $\mathbf{k=}%
\left( k_{x},k_{y}\right) $, we see that $\tilde{\phi}\left( k\right) $ and
the three $\tilde{\phi}_{n}\left( k\right) $ are related as 
\begin{equation}
\begin{tabular}{lll}
$\tilde{\phi}_{n}\left( k\right) =e^{i2\pi \theta _{n}}\tilde{\phi}\left(
k\right) $ & , & $n=1,2,3$ \ ,%
\end{tabular}
\label{fn}
\end{equation}%
with relative phases $\theta _{n}=\mathbf{k}\cdot \mathbf{\delta }_{n}$.
These electronic waves have the same module, $\left\vert \tilde{\phi}%
_{n}\left( k\right) \right\vert ^{2}=\left\vert \tilde{\phi}\left( k\right)
\right\vert ^{2}$; but in general non zero phases; $\theta _{1}\neq \theta
_{2}\neq \theta _{3}$. This means that in the hop of an electron with
momentum $\mathbf{p}=\hbar \mathbf{k}$ from a site $\mathbf{r}_{i}$ to the
nearest at $\mathbf{r}_{i}+\mathbf{\delta }_{n}$, the electronic wave
acquires an extra phase of an amount $\theta _{n}$; but the probability
density at each site is invariant. Demanding the total relative phase to
obey the natural condition, 
\begin{equation}
\theta _{1}+\theta _{2}+\theta _{3}=0\text{ \ },\text{ \ \ }\func{mod}\left(
2\pi \right) ,  \label{te}
\end{equation}%
one ends with the constraint eq(\ref{de}). Let us study two remarkable
consequences of this special conservation law on the $\theta _{n}$ phases by
help of the hidden $SU\left( 3\right) $ symmetry of graphene.\newline
(\textbf{1}) Using eq(\ref{r}), which identifies the relatives $\mathbf{%
\delta }_{n}$\ vectors with the roots $\mathbf{\alpha }_{n}$ of $SU\left(
3\right) $ symmetry, as well as the Lie algebra duality relation 
\begin{equation}
\begin{tabular}{llll}
$\mathbf{\alpha }_{i}\cdot \mathbf{\lambda }_{j}=\delta _{ij}$ & , & $%
i,j=1,2 $ & ,%
\end{tabular}
\label{du}
\end{equation}%
mapping the two simple roots $\mathbf{\alpha }_{1}$, $\mathbf{\alpha }_{2}$
into the $SU\left( 3\right) $ fundamental weights $\mathbf{\lambda }_{1}$, $%
\mathbf{\lambda }_{2}$, we can invert the three equations $\theta _{n}=%
\mathbf{k}\cdot \mathbf{\delta }_{n}$ to get the $\mathbf{p}_{n}\mathbf{=}%
\hbar \mathbf{k}_{n}$ momenta of the electronic waves along the $\mathbf{%
\delta }_{n}$-directions. 
\begin{figure}[tbph]
\begin{center}
\hspace{0cm} \includegraphics[width=14cm]{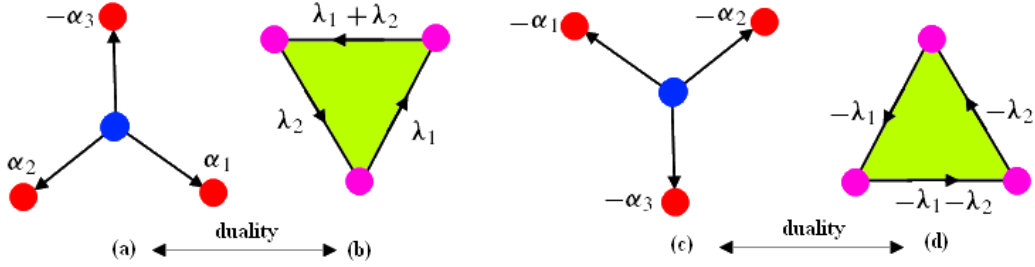}
\end{center}
\par
\vspace{-0.5cm}
\caption{{\protect\small Root/weight duality linking roots and weights }$%
\mathbf{\protect\alpha }_{i}\cdot \mathbf{\protect\delta }_{j}=\protect%
\delta _{ij}$.}
\label{30}
\end{figure}
For the two first $\theta _{n}$'s, that is $n=1,2$, the inverted relations
are nicely obtained by decomposing the 2D wave vector $\mathbf{k}$ along the 
$\mathbf{\lambda }_{1}$ and $\mathbf{\lambda }_{2}$ directions of the dual
lattice to end with the following particular solution,%
\begin{equation}
\begin{tabular}{llll}
$\mathbf{k}_{1}=\theta _{1}\text{ }\mathbf{\lambda }_{1}$ & $,$ & $\mathbf{k}%
_{2}=\theta _{2}\text{ }\mathbf{\lambda }_{2}$ & .%
\end{tabular}
\label{12}
\end{equation}%
More general solutions of type $\mathbf{k}_{1}=\theta _{1}$ $\mathbf{\lambda 
}_{1}+\varkappa _{2}$ $\mathbf{\lambda }_{2}$ and $\mathbf{k}_{2}=\varkappa
_{1}$ $\mathbf{\lambda }_{1}+\theta _{2}$ $\mathbf{\lambda }_{2}$ will be
considered in next subsection. Notice by the way that the 2D vectors $%
\mathbf{\lambda }_{1}$ and $\mathbf{\lambda }_{2}$, interpreted in the
framework of SU$\left( 3\right) $ group theory as the fundamental weights,
are nothing but 
\begin{equation}
\begin{tabular}{llll}
$\mathbf{\lambda }_{1}=\frac{1}{3}\mathbf{a}_{1}$ & $,$ & $\mathbf{\lambda }%
_{2}=\frac{1}{3}\mathbf{a}_{2}$ & ,%
\end{tabular}
\label{la}
\end{equation}%
where $\mathbf{a}_{1,2}=\frac{a}{2}(\pm \sqrt{3},3)$ stand for the
generators of the A- type atoms introduced in the beginning of section 2;
see also fig(\ref{1}). The above relations (\ref{la}) give an other evidence
for the role of the SU$\left( 3\right) $ symmetry in the study of the
electronic properties of graphene.\newline
(\textbf{2}) To get the wave vector from the relation $\theta _{3}=\mathbf{k}%
\cdot \mathbf{\delta }_{3}$, we decompose the 2D vector like $\mathbf{k}%
=q_{1}$ $\mathbf{\lambda }_{1}+q_{2}$ $\mathbf{\lambda }_{2}$; then
substitute $\mathbf{\delta }_{3}=-\mathbf{\alpha }_{1}-\mathbf{\alpha }_{2}$
and use eq(\ref{du}) to end with $\theta _{3}=-q_{1}-q_{2}$. Comparing with
eq(\ref{te}), we find that the wave vector $\mathbf{k}_{3}$ of the
electronic wave along the $\mathbf{\delta }_{3}$- direction reads as follows,%
\begin{equation}
\mathbf{k}_{3}=\theta _{1}\mathbf{\lambda }_{1}+\theta _{2}\mathbf{\lambda }%
_{2}\text{.}  \label{3}
\end{equation}%
Now, combining (\ref{12}) and (\ref{3}), we find that the property $%
\sum_{n}\theta _{n}=0$ describing the conservation law (\ref{te}) of the
total phase of the electron hops to nearest neighbors can be mapped to a
constraint relation on the conservation of total outgoing and incoming
momenta $\hbar \mathbf{k}_{n}$ at each site $\mathbf{r}_{i}$ of the
honeycomb, i.e:%
\begin{equation}
\begin{tabular}{ll}
$\mathbf{k}_{1}+\mathbf{k}_{2}-\mathbf{k}_{3}=\mathbf{0}$ & .%
\end{tabular}
\label{ka}
\end{equation}%
This result is not strange; it may be directly obtained by mapping (\ref{de}%
) to the reciprocal lattice. Below, we study the solutions of this
constraint relation in connection with the hidden $SU\left( 3\right) $
symmetry of the honeycomb.

\subsection{More on the constraint eq(\protect\ref{ka})}

Seen that the relative vectors $\mathbf{\delta }_{n}$ defining the first
nearest neighbors are roots of $SU\left( 3\right) $ as shown by eq(\ref{r}),
a way to deal with the constraint relation (\ref{ka}) is to think about it
as a $SU\left( 3\right) $ group representation relation. This means that the
wave vector in eq(\ref{3}) may be thought of as given by $\mathbf{k}_{1}=\xi
_{1}\mathbf{\Lambda }_{1}$, $\mathbf{k}_{2}=\xi _{2}\mathbf{\Lambda }_{2}$, $%
\mathbf{k}_{3}=\xi _{3}\mathbf{\Lambda }_{3}$ where the $\xi _{n}$'s are
real numbers and $\mathbf{\Lambda }_{1}$, $\mathbf{\Lambda }_{2}$, $\mathbf{%
\Lambda }_{3}$ are three generic weight vectors of $SU\left( 3\right) $. An
interesting situation corresponds to the case where $\xi _{1}=\xi _{2}=\xi
_{3}=\xi $ allowing to turn the constraint eq(\ref{ka}) into a constraint
relation on $SU\left( 3\right) $ weights, 
\begin{equation}
\begin{tabular}{ll}
$\xi \left( \mathbf{\Lambda }_{1}+\mathbf{\Lambda }_{2}-\mathbf{\Lambda }%
_{3}\right) =\mathbf{0}$ & .%
\end{tabular}
\label{lam}
\end{equation}%
This is a remarkable relation which may be motivated by thinking about $%
SU\left( 3\right) $ as a basic symmetry that governs the electronic
properties in graphene. After all, eq(\ref{lam}) is the dual of (\ref{de})
and moreover $\sum_{n=1}^{3}\mathbf{\delta }_{n}=\mathbf{0}$ is itself a $%
SU\left( 3\right) $ condition; see also footnote 1. Under this hypothesis,
and thinking about the $\mathbf{\Lambda }_{n}$'s as \emph{highest} weight
vectors that can be decomposed as,%
\begin{equation}
\begin{tabular}{ll}
$\mathbf{\Lambda }_{n}=p_{n1}\mathbf{\lambda }_{1}+p_{n2}\mathbf{\lambda }%
_{2}$ & ,%
\end{tabular}%
\end{equation}%
we get, after substituting in (\ref{lam}), the following conditions on the $%
p_{ij}$ positive Dynkin integers, 
\begin{equation}
\begin{tabular}{ll}
$\xi \left( p_{11}+p_{21}-p_{31}\right) =0$ & , \\ 
$\xi \left( p_{12}+p_{22}-p_{32}\right) =0$ & .%
\end{tabular}
\label{cl}
\end{equation}%
The simplest solution of these relations corresponds to taking the weight
vectors as 
\begin{equation}
\begin{tabular}{llllll}
$\mathbf{\Lambda }_{1}=\mathbf{\lambda }_{1}$ & , & $\mathbf{\Lambda }_{2}=%
\mathbf{\lambda }_{2}$ & , & $\mathbf{\Lambda }_{3}=\mathbf{\lambda }%
_{adj}=\lambda _{1}+\lambda _{2}$ & ,%
\end{tabular}%
\end{equation}%
in agreement with eqs(\ref{12}-\ref{3}). Particular solutions type $\mathbf{%
\Lambda }=p\mathbf{\lambda }_{1}$ and $\mathbf{\Lambda }^{\prime }=p\mathbf{%
\lambda }_{2}$ with positive integer $p$ are in the same class as $\mathbf{%
\lambda }_{1}$ and $\mathbf{\lambda }_{2}$. From this analysis we learn that
within the $SU\left( 3\right) $ set up, the solutions of (\ref{ka}) have the
following features: \newline
(\textbf{i}) the norm of the wave vector of the $\tilde{\phi}\left(
k_{n}\right) $ wave is $\mathbf{k}_{n}^{2}=\frac{2}{3}\left(
p_{n1}^{2}+p_{n2}^{2}+p_{n1}p_{n2}\right) \xi ^{2}$,\newline
(\textbf{ii}) the phases of the waves $\tilde{\phi}_{n}\left(
k_{x},k_{y}\right) $, defined by $\varphi _{n}=\arctan \left( \frac{k_{ny}}{%
k_{nx}}\right) $, are quantized as%
\begin{equation}
\begin{tabular}{ll}
$\frac{k_{ny}}{k_{nx}}=\frac{\left( p_{n1}+p_{n2}\right) }{p_{n1}-p_{n2}}%
\sqrt{3}$ & $.$%
\end{tabular}%
\end{equation}%
(\textbf{iii}) the Fourier waves $\tilde{\phi}\left( k_{x},k_{y}\right) $
may be interpreted as 1D field multiplets transforming into $SU\left(
3\right) $ highest weight representations as, 
\begin{equation}
\begin{tabular}{ll}
$\tilde{\phi}\left( k_{x},k_{y}\right) =\tilde{\phi}_{\Lambda }\left( \xi
\right) =\tilde{\phi}_{\left( p,q\right) }\left( \xi \right) $ & ,%
\end{tabular}%
\end{equation}%
with dimension%
\begin{equation}
\begin{tabular}{ll}
$\frac{1}{2}\left( 1+p\right) \left( 1+q\right) \left( 2+p+q\right) $ & .%
\end{tabular}%
\end{equation}%
In this $SU\left( 3\right) $ picture, the physics describing the electron
hops between the nearest carbons is completely captured by $SU\left(
3\right) $ highest state representations. For instance, taking $\Lambda
_{1}\equiv $ \b{3} and $\mathbf{\Lambda }_{2}\equiv \bar{3}$ as in fig(\ref%
{4}), and using the tensor product decomposition 
\begin{equation}
\begin{tabular}{ll}
$\text{\b{3}}\otimes \bar{3}=9=\text{\b{8}}\oplus \text{\b{1}}$ & ,%
\end{tabular}%
\end{equation}%
the Fourier waves propagating between the nearest sites are $\tilde{\phi}%
_{\left( 1,0\right) }$, $\tilde{\phi}_{\left( 0,1\right) }$, $\tilde{\phi}%
_{\left( 1,1\right) }$ and $\tilde{\phi}_{\left( 0,0\right) }$. The zero
mode $\tilde{\phi}_{\left( 0,0\right) }$ describes the state where the
electron doesn't hop. 
\begin{figure}[tbph]
\begin{center}
\hspace{0cm} \includegraphics[width=8cm]{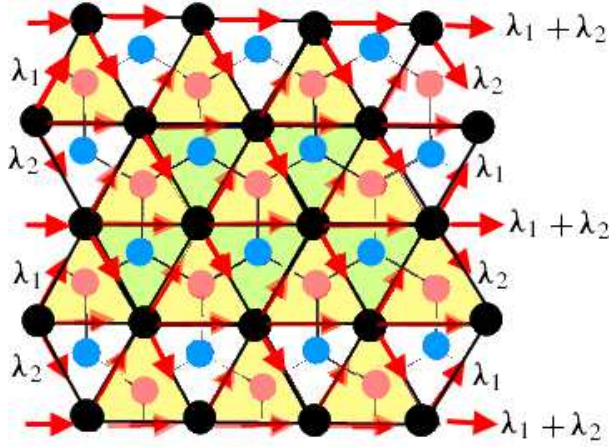}
\end{center}
\par
\vspace{-0.5 cm}
\caption{{\protect\small Real and dual lattices represented simultaneously.
Total probability current density and total momenta of the electronic waves
are conserved at each site of the honeycomb.}}
\label{4}
\end{figure}
Notice that each of these $\phi _{\Lambda }\left( \xi \right) $ fields hide
sub-modes $\left( \phi _{\mathbf{\lambda }}\right) ^{\mu }$ associated with
the various states of the representations. For the example of fig(\ref{4}),
we have: 
\begin{equation}
\begin{tabular}{llllllll}
$\phi _{\mathbf{\lambda }}$ & $\equiv $ & $\phi _{\mathbf{\lambda }%
}^{\lambda }$ & , & $\phi _{\mathbf{\lambda }}^{\lambda -\alpha _{%
{\scriptsize 1}}}$ & , & $\phi _{\mathbf{\lambda }}^{\lambda -\psi }$ & , \\ 
$\phi _{\mathbf{\bar{\lambda}}}$ & $\equiv $ & $\phi _{\mathbf{\bar{\lambda}}%
}^{\mathbf{\bar{\lambda}}}$ & , & $\phi _{\mathbf{\bar{\lambda}}}^{\mathbf{%
\bar{\lambda}}-\alpha _{{\scriptsize 2}}}$ & , & $\phi _{\mathbf{\bar{\lambda%
}}}^{\mathbf{\bar{\lambda}}-\psi }$ & , \\ 
$\phi _{\mathbf{\lambda }_{^{{\scriptsize adj}}}}$ & $\equiv $ & $\phi
_{\psi }^{\pm \psi }$ & , & \ $\phi _{\psi }^{\pm \left( \psi -\alpha _{%
{\scriptsize 1,2}}\right) }$ & , & $\phi _{\psi }^{\pm 0}$ & ,%
\end{tabular}%
\end{equation}%
with $\mathbf{\lambda }^{{\scriptsize adj}}=\psi $. This degeneracy can be
lifted by breaking down the hidden $SU\left( 3\right) $ symmetry of the
graphene. This may be achieved by implementing defects in the honeycomb\
that violate eq(\ref{cl}).

\section{Energy dispersion and the hidden symmetries}

The energy $h_{i}$ describing the hopping of a pi-electron from the site $%
\mathbf{r}_{i}$ to its three nearest neighbors at $\mathbf{r}_{i}+\mathbf{%
\delta }_{n}$ is nicely represented by the tight binding hamiltonian\textrm{%
\ \cite{16} }whose total form reads as $H=-t\sum_{i}h_{i}$,%
\begin{equation}
h_{i}=\sum_{n=0}^{2}a\left( \mathbf{r}_{i}\right) b^{\dagger }\left( \mathbf{%
r}_{i}+\mathbf{\alpha }_{n}\right) +hc,  \label{t}
\end{equation}%
where $a_{i},$ $b_{j}$, $a_{i}^{\dagger }$, $b_{j}^{\dagger }$ are fermionic
annihilation and creation oscillators and $t\simeq 2.8eV$ the hopping
energy. With this hamiltonian H, one learns much about the electronic band
structure of graphene. However to get more insight about the hidden
symmetries of the honeycomb, it is interesting to express H in terms of the $%
F_{i}^{\pm \alpha _{n}}$ steps operators generating SU$\left( 2\right) $
sub-symmetries inside SU$\left( 3\right) $. To do so, we start from the wave
functions $\phi _{A}\left( \mathbf{r}_{i}\right) \equiv <\mathbf{r}|\phi
_{i}>$ and $\phi _{B}\left( \mathbf{r}_{i}+\mathbf{\alpha }_{n}\right)
\equiv <\mathbf{r}|\phi _{i}^{\mathbf{\alpha }_{n}}>$ associated with a
fixed A-type atom and its nearest B- type neighbors. Then use the structure
of the honeycomb (fig(\ref{1})) to write down the action of the $F_{i}^{\pm
\alpha _{n}}$'s generating the electron hopping. At each $\mathbf{r}_{i}$ of
the sublattice A, we have 
\begin{equation}
\begin{tabular}{lllll}
$F_{i}^{+\mathbf{\alpha }_{n}}\left\vert \phi _{i}\right\rangle $ & $%
=\left\vert \phi _{i}^{\mathbf{\alpha }_{n}}\right\rangle $ & , & $F_{i}^{+%
\mathbf{\alpha }_{n}}\left\vert \phi _{i}^{\mathbf{\alpha }%
_{n}}\right\rangle =0$ & , \\ 
$F_{i}^{-\mathbf{\alpha }_{n}}\left\vert \phi _{i}^{\mathbf{\alpha }%
_{n}}\right\rangle $ & $=\left\vert \phi _{i}\right\rangle $ & , & $F_{i}^{-%
\mathbf{\alpha }_{n}}\left\vert \phi _{i}\right\rangle =0$ & ,%
\end{tabular}%
\end{equation}%
from which we read the following relations, 
\begin{equation}
\begin{tabular}{lllll}
$\left[ F_{i}^{+\mathbf{\alpha }_{n}},F_{i}^{-\mathbf{\alpha }_{n}}\right] $
& $=h_{i}^{\mathbf{\alpha }_{n}}$ & , & $\left[ h_{i}^{\mathbf{\alpha }%
_{n}},F_{i}^{\pm \mathbf{\alpha }_{n}}\right] =\pm F_{i}^{\pm \mathbf{\alpha 
}_{n}}$ & , \\ 
$\left\{ F_{i}^{\pm \mathbf{\alpha }_{n}},F_{i}^{\pm \mathbf{\alpha }%
_{n}}\right\} $ & $=0$ & , & $\left\{ F_{i}^{\mathbf{\alpha }_{n}},F_{i}^{-%
\mathbf{\alpha }_{n}}\right\} =J_{i}^{\mathbf{\alpha }_{n}}$ & .%
\end{tabular}
\label{cr}
\end{equation}%
with $J_{i}^{\mathbf{\alpha }_{n}}$ a commuting central element $\left[
F_{i}^{\pm \mathbf{\alpha }_{n}},J_{i}^{\mathbf{\alpha }_{n}}\right] =0$.
The commutation relations tell us that \emph{locally} each set $\left(
h_{i}^{\mathbf{\alpha }_{n}},F_{i}^{\pm \mathbf{\alpha }_{n}}\right) $
generate an $SU\left( 2\right) $ group along the $\mathbf{\alpha }_{n}$%
-direction in the hidden SU$\left( 3\right) $ symmetry. The anti-commutation
relations, which read also like $\left( F_{i}^{\pm \mathbf{\alpha }%
_{n}}\right) ^{2}=0$, requires $F_{i}^{\pm \alpha _{n}}$ to be in the
isospin $\frac{1}{2}$ representation; that is $2\times 2$ matrices linking
the two sublattices A and B of the honeycomb. The fermionic realization (\ref%
{t}) is a representation of eqs(\ref{cr}) where the $F_{i}^{\pm \mathbf{%
\alpha }_{n}}$'s are solved as%
\begin{equation}
\begin{tabular}{llll}
$F_{i}^{+}=a_{i}b_{i}^{\dagger }$ & , & $h_{i}=b_{i}b_{i}^{\dagger
}-a_{i}a_{i}^{\dagger }$ & , \\ 
$F_{i}^{-}=a_{i}^{\dagger }b_{i}$ & , & $J_{i}=b_{i}b_{i}^{\dagger
}+a_{i}a_{i}^{\dagger }$ & .%
\end{tabular}%
\end{equation}%
In terms of the globally defined operators $F^{\pm \mathbf{\alpha }%
_{n}}=\sum_{i}F_{i}^{\pm \mathbf{\alpha }_{n}}$, the hamiltonian $H$ takes
the simple form,%
\begin{equation}
H=-t\sum_{n=0}^{2}\left( F^{+\mathbf{\alpha }_{n}}+F^{-\mathbf{\alpha }%
_{n}}\right) ,  \label{h}
\end{equation}%
where we have used $\mathbf{\alpha }_{0}=-\mathbf{\alpha }_{3}$. Besides
hermiticity, $H$ has two special features that we want comment: (\textbf{i})
H is not invariant under the $SU\left( 2\right) $ symmetries since along
with (\ref{h}) we also have the cousin operators 
\begin{equation}
\begin{tabular}{ll}
$L=-i\vartheta \sum\limits_{n=0}^{2}\left( F^{+\mathbf{\alpha }_{n}}-F^{-%
\mathbf{\alpha }_{n}}\right) $, & $M=\sum\limits_{n=0}^{2}h^{\mathbf{\alpha }%
_{n}}$,%
\end{tabular}
\label{f}
\end{equation}%
obeying the commutation relations 
\begin{equation}
\begin{tabular}{lll}
$\left[ \frac{H}{t},\frac{L}{i\vartheta }\right] =M$, & $\left[ M,\frac{H}{t}%
\right] =\frac{L}{i\vartheta }$, & $\left[ M,\frac{L}{i\vartheta }\right] =%
\frac{H}{t}$.%
\end{tabular}%
\end{equation}%
The real number $\vartheta $ in (\ref{f}) may be interpreted in terms of
coupling to a constant external magnetic field. (\textbf{ii}) H is not a
positive definite operator in the sense that its energy spectrum has two
signs; a region with positive energy describing the conduction band and a
region with negative energy associated with holes.\newline
Performing the Fourier transform of the step operators $F^{+\mathbf{\alpha }%
_{n}}=\sum e^{i2\pi \mathbf{k}\cdot \mathbf{\alpha }_{n}}G_{k}$ and putting
back into the hamiltonian, we can put $H$ in various forms; in particular
like 
\begin{equation}
H=-t\sum \left( \psi _{k}G_{k}+\bar{\psi}_{k}\bar{G}_{k}\right) ,
\end{equation}%
with $\psi _{k}=e^{i2\pi \mathbf{k}\cdot \mathbf{\alpha }_{1}}+e^{i2\pi 
\mathbf{k}\cdot \mathbf{\alpha }_{2}}+e^{-i2\pi \mathbf{k}\cdot \mathbf{%
\alpha }_{3}}$. Setting $Q=e^{i2\pi \xi }$ and $\mathbf{k}=\xi \Lambda $
with $\Lambda =p_{1}\lambda _{1}+p_{2}\lambda _{2}$, we can bring this
hamiltonian to $-t\sum_{\xi ,\Lambda }H_{\Lambda }\left( \xi \right) $ with 
\begin{equation}
H_{\Lambda }=\psi _{\Lambda }^{su\left( {\small 3}\right) }G_{\Lambda }+\bar{%
\psi}_{\Lambda }^{su\left( {\small 3}\right) }\bar{G}_{\Lambda },
\end{equation}%
and $\psi _{\Lambda }^{su\left( {\small 3}\right) }=\left[ Q^{\mathbf{%
\Lambda }.\mathbf{\alpha }_{1}}+Q^{\mathbf{\Lambda }.\mathbf{\alpha }%
_{2}}+Q^{-\mathbf{\Lambda }.\mathbf{\alpha }_{1}-\mathbf{\Lambda }.\mathbf{%
\alpha }_{2}}\right] .$ To get the wave vectors $\mathbf{K}_{F}=\xi
_{F}\Lambda _{F}$ at the Fermi level, one has to solve the zero energy
condition $\psi _{\Lambda }^{su\left( {\small 3}\right) }\left( \xi \right)
=0$ whose solutions are given by the cubic root of unity $\left(
1+Q+Q^{2}\right) =0$. They are generated by $\xi _{F}=\frac{1}{3}$ times the
fundamental weights of the dual lattice; i.e $\mathbf{K}_{F}=\frac{1}{3}%
\mathbf{\lambda }_{1}$ and $\mathbf{K}_{F}^{\prime }=\frac{1}{3}\mathbf{%
\lambda }_{2}$ modulo translations. Moreover setting 
\begin{equation}
\begin{tabular}{llll}
$G_{k}=e^{iT_{G}}\mathcal{P}_{k}$ & , & $\psi _{k}=\left\vert \psi
_{k}\right\vert e^{i\vartheta _{\psi }}$ & ,%
\end{tabular}%
\end{equation}%
where $\mathcal{P}_{k}$ is positive definite and $T_{G}$ hermitian; then
substituting in (\ref{h}), we get%
\begin{equation}
H=\sum_{k}\varepsilon _{k}\mathcal{N}_{k}-\sum_{k}\varepsilon _{k}\mathcal{M}%
_{k}\text{ },
\end{equation}%
with $\varepsilon _{k}=\left\vert t\psi _{k}\right\vert $ and%
\begin{equation}
\begin{tabular}{ll}
$\mathcal{M}_{k}=2\cos ^{2}\left( \frac{\vartheta _{\psi }+T_{G}}{2}\right) 
\mathcal{P}_{k}$ & , \\ 
$\mathcal{N}_{k}=2\sin ^{2}\left( \frac{\vartheta _{\psi }+T_{G}}{2}\right) 
\mathcal{P}_{k}$ & .%
\end{tabular}
\label{nm}
\end{equation}%
We end this section by first noting that using the fermionic realization, eq(%
\ref{nm}) gets a simple interpretation in terms of electron and hole number
operators $e_{k}^{\dagger }e_{k}$ and $h_{k}^{\dagger }h_{k}$. Second, the
group theoretical approach developed in this study may be used to deal with
graphene multilayers. In the case of graphene bilayer, one expects
symmetries of type $SU\left( 3\right) \times SU\left( 2\right) \times
SU\left( 3\right) $ with each $SU\left( 3\right) $ factor as before and
where the $SU\left( 2\right) $ term refers to transitions between the two
layers.

\section{Conclusion}

In this paper, we have shown that 2D graphene has a remarkable hidden $%
SU\left( 3\right) $ symmetry that allow to classify the propagating Fourier
waves $\tilde{\phi}\left( k_{x},k_{y}\right) $ in terms of 1D highest weight
field multiplets $\tilde{\phi}_{\Lambda }\left( \xi \right) $. Conservation
of total incoming and outgoing momenta at each lattice site translates into
triplets of $SU\left( 3\right) $ HWRs constrained by eq(\ref{cl}) and whose
basic one is $\left( 3,\bar{3},8\right) $ satisfying $3\otimes \bar{3}%
=8\oplus 1$. We have also shown that, from $SU\left( 3\right) $ view, $\tan
\varphi =\frac{k_{y}}{k_{x}}$ is quantized as $\frac{\left( p+q\right) }{%
\left( p-q\right) }\sqrt{3}$ and moreover the hamiltonian of the tight
binding model is the sum of the step operators of the three $SU\left(
2\right) $ sub-symmetries of the hidden $SU\left( 3\right) $ invariance.
This connection with Lie algebra teaches us that graphene may be thought of
as the second element of class of theoretical systems involving higher rank
symmetries \textrm{\cite{15,16}}. The first element has a hidden $SU\left(
2\right) $ and energy spectrum $\varepsilon _{\Lambda }^{2}\left( \xi
\right) =t^{2}\psi _{\Lambda }^{su\left( {\small 2}\right) }\bar{\psi}%
_{\Lambda }^{su\left( {\small 2}\right) }$; this should correspond to the
linear poly-acetylene chain with hamiltonian $H^{su\left( 2\right)
}=-t\sum_{i}h_{i}^{su\left( 2\right) }$ and,%
\begin{equation}
h_{i}^{su\left( 2\right) }=a\left( \mathbf{r}_{i}\right) b^{\dagger }\left( 
\mathbf{r}_{i}+\mathbf{\alpha }\right) +a\left( \mathbf{r}_{i}\right)
b^{\dagger }\left( \mathbf{r}_{i}-\mathbf{\alpha }\right) +hc,
\end{equation}%
where $\alpha $ stands for the $SU\left( 2\right) $ root. It is also
interesting to note the connection between $\sum_{n=1}^{3}\mathbf{\delta }%
_{n}=\mathbf{0}$ and the imaginary root of affine $S\hat{U}\left( 3\right) $
suggesting that ideal graphene could have a richer hidden symmetry
containing $SU\left( 3\right) $ as the zero mode. If this is the case,
graphene would also exhibit a hidden 2D conformal structure underlying the
honeycomb lattice and capturing information on eventual critical behaviors
of electronic correlations along the line of \textrm{\cite{17,15}}. This
issue is understudy; it will be developed elsewhere.

\end{document}